# Nearly Complete Segregation of Submerged Grains in a Rotating Drum

Yu Chen[1], Deheng Wei[1,2,†], Si Suo[3], Mingrui Dong[1], Yixiang Gan[1,*]

[1] School of Civil Engineering, The University of Sydney, Sydney, Australia
[2] Key Laboratory of Ministry of Education on Safe Mining of Deep Mines, School of Resources and Civil Engineering, Northeastern University, Shenyang, 110819, China.
[3] Department of Civil and Environmental Engineering, Imperial College London, London, the UK

Density-driven segregations, extensively studied in a simple rotating drum, are enriched with a wide range of underlying physics. Diverse symmetrical segregation patterns formed by mixing two types of dry mono-sized grains have been revealed due to variations in heavy and light grain densities, $\rho_h$ and $\rho_l$, and rotating speeds, $\omega$. We engender experimentally a nearly complete segregation, not occurring in dry conditions of the same $\rho_h$, $\rho_l$, and $\omega$, in submerged states. Further, based on the experiment-validated simulations, using coupled computational fluid dynamics and discrete element method, it is found the mixing index can be well predicted over a wide parameter space in the effective density ratio, $\mathcal{D} = (\rho_h - \rho_f)/(\rho_l - \rho_f)$ with $\rho_f$ being the fluid density. Specifically, with increasing $\mathcal{D}$ well-mixed states transit to fully-segregated states with a rising number of vortices and severer asymmetrical patterns. When the global Reynolds number $Re_g$ is enlarged, the vortex area of heavy particles shrinks for lower $\mathcal{D}$, while the area of light particles gradually saturates; meanwhile, for higher $\mathcal{D}$ a new vortex with a continuously expanded area can be encountered in the light particle zone. These results improve our understanding of segregation transitions especially in submerged granular systems and shed new light on various science and engineering practices.

† Email address for correspondence: deheng.wei@sydney.edu.au
* Email address for correspondence: yixiang.gan@sydney.edu.au

## 1. Introduction

Understanding granular segregation is of vital importance in many geophysical problems and industrial applications, ranging from sediment transport (Ferdowsi et al., 2017; Iverson, 1997) to mixing processes in pharmaceutical and wood pulp engineering (Badar and Farooqi, 2012; Sommier et al., 2001). Granular materials often exhibit spontaneous segregation based on differences in size, shape, and density, leading to distinct phase separations (Umbanhowar et al., 2019). Proceeding from the minimal case of mixing two types of spherical grains in a thin rotating cylindrical container, the chaotic mixing and segregation in dry granular flows is well studied on density-dispersity in mono-sized particles (D-system) and size-dispersity in grains of the same density (S-system), where the types of grains in each system are identical. In the former, 'central core' patterns of heavy grains sunk into light grains could be formed in the symmetrical mixtures (Barker et al., 2020; Khakhar et al., 2003; Zuriguel et al., 2009), due to the buoyancy-driven mechanism (Duan et al., 2020; Gray, 2018; Vallance and Savage, 2000). Such a symmetry, induced by high grain density, $\mathfrak{D} = \rho_h/\rho_l$, might hinder the postmortem de-mixing by potential re-mixing. This can be alleviated by rotating the two binary density grains in submerged conditions (Liao, 2018), wherein it is observed that the "central core" moves downward along the incline slope. Inspired by the effects of interstitial fluids on D-system, we aim to tackle a nearly complete segregation with the "central core" clustering near the drum boundary.

It is in this light surprising that so little attention has been paid to quantifying the segregation efficiency of wet rotating drums. One of the main reasons for this is that the particle-fluid interaction force is inherently difficult to measure, unlike particle-particle contact force; the intriguing, yet dominant parts of it (i.e., drag and lift forces) are scaled with the root of the relative velocity (Sun et al., 2010). To date, the velocity-dependent drag and lift coefficients can be only determined indirectly, by more or less empirically fittings (Vergara et al., 2024). Considering the high-velocity gradient in a drum to attain high grain segregation intensity, the configuration is too experimentally complicated to delineate quantitative influences of each fluid properties. The numerical modelling is therefore an effective candidate to isolate the effect of focused factors, such as the effective density ratio, $\mathcal{D} = (\rho_h - \rho_f)/(\rho_l - \rho_f)$, in this study, from other dominants, on critical metrics quantifying grain segregation.

Numerous experimental and numerical investigations have been conducted on the dry granular segregation in D-systems. The mixing efficiency of binary grain mixtures has been frequently measured by the mixing index, firstly proposed by Lacey (1954). By tracking trajectories of the heavier particles in a quasi-2D drum experimentally, Ristow (1994) quantified the segregation rate and its dependence on the dry particle density ratio, $\mathfrak{D}$, at a cascading regime. Khakhar et al. (1997) further demonstrated that the composition profiles of heavy and light grains remained independent of the level of particle filling. Using numerical simulations of distinct element method (DEM), the settling dynamics of a single high-density particle within a granular mixture illustrates that the motion of the high-density tracer is characterised by the 'segregation force', combining with grain-grain buoyancy and drag forces (Tripathi and Khakhar, 2013).

Buoyancy-driven segregation of dry grains often engenders the formation of distinct layers, observed in both vibrated granular mixtures (Shi et al., 2007; Tai et al., 2010) and rotating drums (Liao et al., 2024; Pereira et al., 2010). Lighter particles migrate towards the periphery and form a segregated layer, leading to stratified configurations. These regimes include rolling and cascading regimes, with an elliptical vortex observed at the core of the flow and highly overlapping with the convection roll cores of heavier particulates (Chen et al., 2024). Much attention has been paid to identifying the only vortex in dry D-system by tracking streamlines to determine its location and area (Arntz et al., 2014; Romanò et al., 2017; Tang et al., 2024). How the vertex in submerged D-system evolutes by the combined effects of $\mathcal{D}$ and $\omega$ deserves, at least, equal attention.

In this work, we aim to investigate the nearly complete segregation behaviour of submerged binary-mixed grains in a rotating drum. We employ a combined experimental and numerical method to systematically explore how the effective density ratio and rotating speed influence segregation patterns. We identify four different states of mixing, ranging from homogenous mixing to the development of distinct segregated regions by multiple vortices formation. Corresponding mixing states transition from well-mixed to nearly complete segregation state is influenced by the effective density ratio, irrespective of fluid density, and characterised by the mixing index; further transitions in nearly complete segregated stages are highly sensitive to both effective density ratio and rotating speed. By gaining a deeper understanding of these segregation states, this research provides a clearer picture of nearly complete segregation processes in a submerged D-system within a rotating drum.

## 2. Methodology

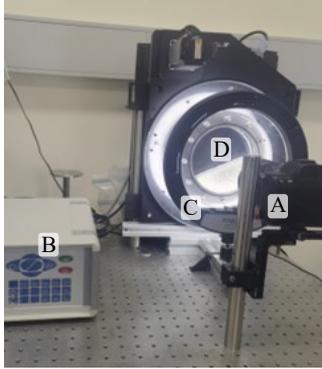
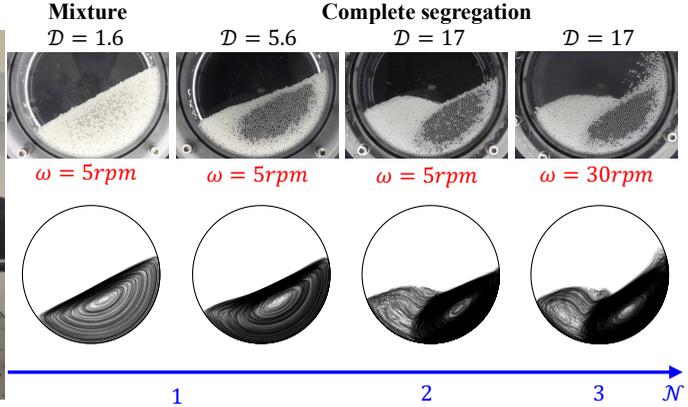

FIGURE 1. (a) Experimental setup: A, B, C, and D are the camera, rotating controller, light system, and counterclockwise-rotated thin cylindrical container, respectively. (b) Flow patterns observed at various $\mathcal{D} = (\rho_h - \rho_f)/(\rho_l - \rho_f)$ and $\omega$, and $\mathcal{N}$ denotes the number of vortices. The top row displays the experimental observations, while the bottom row illustrates the particle trajectories derived from the corresponding numerical simulations.

*2.1 Experimental configuration*

Fig. 1(a) shows our experimental setup, consisting of a drum container with an inner diameter of 150 mm and a thickness of 30 mm controlled by a rotator proving rotating speed within [5, 60] rpm. The mixtures were composed of binary spherical particles with different densities selected from three materials: polyoxymethylene (POM, $\rho_s = 1{,}382$ kg/m$^3$), glass beads ($\rho_s = 2{,}363$ kg/m$^3$), and stainless steel ($\rho_s = 7{,}803$ kg/m$^3$). All of them have the same nominal diameter, $d_p$, of 2 mm $\pm$ 3%. The initial mixtures were prepared with an equal number of 14,274 of particles of the two selected types. The corresponding packing fraction was approximately 0.62; and the filling ratio, defined as the ratio of the total filled volume to the drum volume, was about 0.356. In the experiment, the submerged condition referred to grains being rotating in a drum fully filled by purified water whose density $\rho_w = 1{,}000$ kg/m$^3$. All tests were conducted in a laboratory maintained at the room temperature of 22°C and the atmosphere pressure of 1atm, resulting in the water viscosity of 0.001 Pa·s. The flow patterns were recorded until the steady flow state was achieved. Fig. 1(b) shows flow patterns at steady state with various conditions. The experimental procedures are designed to ensure repeatability, with all critical parameters—particle size distribution, packing fraction, drum filling ratio, and rotating speed—carefully controlled to minimise variability between trials. Further details of the experiment can be seen in our previous work (Chen et al., 2024).

*2.2 Governing equations and numerical scheme*

An unresolved 3D CFD-DEM approach (Tsuji et al., 1993) is employed to simulate the dynamics in the D-system, implemented using the open-source CFDEMPROJECT library. The fluid phase is treated as a continuum and modelled in an Eulerian framework (Zhou et al., 2010) using local-averaged Navier-Stokes equations:

$$\frac{\partial}{\partial t}(\varepsilon_f \rho_f) + \nabla \cdot (\varepsilon_f \rho_f \boldsymbol{u}_f) = 0, \tag{1}$$

$$\frac{\partial}{\partial t}(\varepsilon_f \rho_f \boldsymbol{u}_f) + \nabla(\varepsilon_f \rho_f \boldsymbol{u}_f \boldsymbol{u}_f) = -\varepsilon_f \nabla p_f + \nabla \cdot (\varepsilon_f \mathbf{T}_f) + \varepsilon_f \rho_f \boldsymbol{g} - \mathbf{F}_{pf}, \tag{2}$$

where $\varepsilon_f$ is the local fluid fraction, $\rho_f$ the fluid density, $\boldsymbol{u}_f$ the fluid velocity, and the stress tensor $\mathbf{T}_f = \mu_f(\nabla \boldsymbol{u}_f + \nabla \boldsymbol{u}_f^T)$. Other items on the right side of Eq. (2) denote the pressure gradient, gravitational force, and particle-fluid interaction force, respectively.

The particle dynamics are solved using a DEM framework, in which both translational and rotational motions are accounted. These distinct motions are governed by the principles of Newton's second law and the angular momentum equation:

$$m_i \frac{d\boldsymbol{v}_i}{dt} = m_i \boldsymbol{g} + \boldsymbol{f}_{c,i} + \boldsymbol{f}_{pf,i}, \tag{3}$$

$$I_i \frac{d\boldsymbol{\omega}_i}{dt} = \boldsymbol{M}_i, \tag{4}$$

where $m_i$, $\boldsymbol{v}_i$, $I_i$ and $\boldsymbol{\omega}_i$ denote the mass, translational velocity, moment of inertia and angular velocity of the $i$-th particle, respectively. For the inter-particle contact, $\boldsymbol{f}_{c,i}$ is composed of two parts, i.e., the normal force solved by the Hertzian contact model, and the tangential one governed by the finite Coulomb friction model and characterised by the friction coefficient, $\mu_s$. For the particle rotation, we consider a rolling resistance for calculating the total moment $\boldsymbol{M}_i$, characterised by the rolling friction coefficient, $\mu_r$. The collision energy loss by the inelasticity is considered through a restitution coefficient, $e$, in both normal and tangential directions (Brilliantov et al., 1996; Cundall and Strack, 1979). For the particle-fluid interaction, $\boldsymbol{f}_{pf,i}$ includes both the Archimedes buoyance force $\boldsymbol{f}_{b,i} = V_i(-\nabla p_f + \nabla \cdot \mathbf{T}_f)$ and drag force $\boldsymbol{f}_{d,i} = \beta V_i(\boldsymbol{u}_{f,i} - \boldsymbol{v}_i)/(\varepsilon_f - \varepsilon_f^2)$. Herein, $V_i$ is the volume of particle $i$ and $\beta$ is the inter-phase momentum exchange coefficient determined by the Gidaspow drag model (Gidaspow, 1994),

$$\beta = \begin{cases} \dfrac{150(1-\varepsilon_f)\boldsymbol{u}_f}{\varepsilon_f d_p^2} + 1.75\dfrac{(1-\varepsilon_f)\rho_f|\boldsymbol{u}_f - \boldsymbol{v}_p|}{d_p} & \varepsilon_f \leq 0.8 \\ \dfrac{3}{4}\dfrac{\varepsilon_f(1-\varepsilon_g)\rho_f|\boldsymbol{u}_f - \boldsymbol{v}_p|}{d_p}C_D\varepsilon_f^{-2.65} & \varepsilon_f > 0.8 \end{cases}, \tag{5}$$

$$C_D = \begin{cases} \dfrac{24}{Re}(1 + 0.15Re_p^{0.687}) & Re_p \leq 1000 \\ 0.44 & Re_p > 1000 \end{cases}, \tag{6}$$

where $C_D$ is the drag coefficient dependent on the particle Reynolds number, $Re_p = \rho_f \varepsilon_f d_p|\boldsymbol{u}_f - \boldsymbol{v}_p|/\mu_f$. Meanwhile, to characterize the overall flow regime in the rotating drum, a global Reynolds number, $Re_g = \rho_f \omega r d_p/\mu_f$, is defined with $r$ being the radius of the cylindrical drum.

For the numerical setup, the drum geometry, particle diameters, and drum filling ratio are consistent with the experimental configurations. The computational cell size in CFD is set to $2d_p$ and the time steps for CFD and DEM are $10^{-3}$ s and $10^{-4}$ s, respectively, verified by sensitivity studies. To calibrate the model parameters regarding friction and collision, a benchmark simulation is conducted based on the mono-dispersed granular flows. The angles of repose at different rotating speed are measured under submerged conditions for three different particle densities. All other parameters, like fluid viscosity ($\mu_f = 0.001 \text{Pa} \cdot \text{s}$), and Young's modulus and Poisson's ratio of grains, are kept consistent with the experiment. Consequently, $\mu_s = 0.3$, $\mu_r = 0.0002$, and $e = 0.5$.

To quantify the segregation between binary mixtures, the Lacey mixing index (Lacey, 1954) is applied:

$$M = \frac{(\sigma^2 - \sigma_0^2)}{(\sigma_m^2 - \sigma_0^2)} \in [0,1], \quad (7)$$

with $\sigma^2 = \sum_{i=1}^{N} \frac{(\phi_i - \phi_m)^2}{N-1}$, where $N$ is the number of identical cells occupied by all grains, and $\phi_i$ and $\phi_m$ are the local and the mean concentration of tracer grains, respectively; $\sigma_0^2 = \phi_m \cdot (1 - \phi_m)$ is the theoretical upper limit, while $\sigma_m^2 = \frac{\sigma_0^2}{n}$, with $n$ being the grain number in the $i$-th cell, is the possible minimum value. As shown in the inset of Fig. 2(a) on the mixing index, the numerical and experiment results ($M_s$ and $M_e$), have a maximum relative error is 2.8%. Notably, due to the difficulty in quantifying the 3D experimental postmortem $M$, both $M_s$ and $M_e$ are measured from visible layers close to the cylinder boundaries. In the following, we extend the range of the effective density ratio $\mathcal{D}$ by varying fluid density $\rho_f \in [0.2\rho_w, \rho_w]$ and particle density $\rho_s \in [1.400\rho_w, 9.743\rho_w]$.

## 3. Results and discussions

3.1 *Effect of effective density ratio*

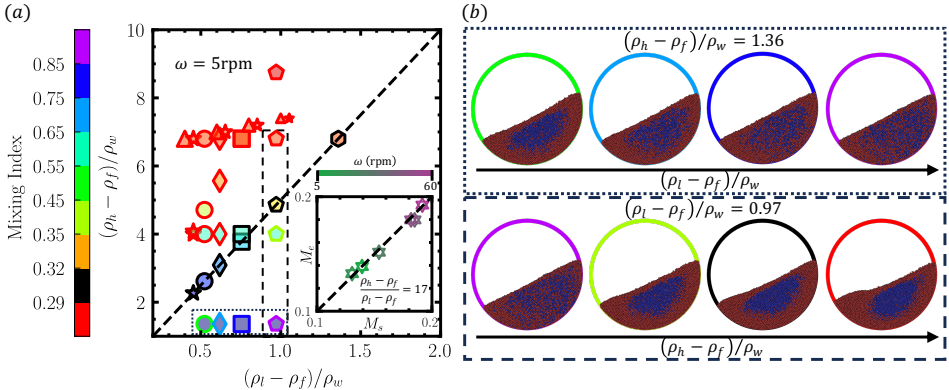

FIGURE 2. (a) Phase diagram illustrating the mixing index ($M$), as in Eq. (7), across various combinations of heavy $\rho_h$ and light $\rho_l$ particle densities, as well as fluid densities $\rho_f$, at a fixed rotating speed, $\omega = 5$ rpm. The black dashed line corresponds to $(\rho_h - \rho_f)/(\rho_l - \rho_f) = 5$. The marker edge colour represents the values of $M$, while filled colour, shape, and size of the marker indicate values of $\rho_h$, $\rho_l$, and $\rho_f$, respectively, consistent with those in Fig. 3(a). The inset compares the numerical ($M_s$) and experimental ($M_e$) mixing indices for submerged steel and POM mixtures with various rotating speeds. (b) Visualisations of specific steady mixing states in (a), in which the outer ring colour indicates $M$ value.

Experimental observations demonstrate four distinctive mixing patterns of submerged D-systems, as shown in the top of Fig. 1(b). The well-mixed state is pertinent to that of dry system. However, the segregation patterns are totally different; that is, the net symmetry of the segregation profiles diminishes as seen from the eccentricity of heavy grains, which is enhanced with $\omega$ increasing. Meanwhile, the heavy grains become more clustering and closer to the boundary of the drum, leading to the nearly complete segregation. Notably, for the POM-glass mixture ($\mathcal{D} = 1.6$), no matter how high $\omega$ is, no visible segregation was encountered, which is against the observations in the POM-steel ($\mathcal{D} = 17$) and the steel-glass ($\mathcal{D} = 5$) mixtures. To gain deeper insights, especially for the grain scale, such as contact number distributions, into the segregation transition, extra CFD-DEM simulations are conducted to investigate density effects.

Because the segregation patterns are dominated by the gravity and buoyancy forces proportional to the density differences, Fig. 2 shows a phase diagram for $M$ influenced by $\rho_h - \rho_f$ and $\rho_l - \rho_f$. We define $M \geq 0.32$ as the mixed state, and $M < 0.32$ shows segregation with two distinct modes. The regime of $0.29 < M < 0.32$ has been explored previously as the central core segregation mode (Liao et al., 2024; Zuriguel et al., 2009), whilst $M \leq 0.29$ breaks the symmetry from the previous mode and transits towards nearly complete segregation. As illustrated by data points in the dashed and dotted rectangles for the same $\omega$ and $\rho_f$ in Fig. 2(a), of which the steady-state configurations are provided in Fig. 2 (b), $M$ is not uniquely dependent on $\rho_h$ or $\rho_l$, but on the combined effect of the two with the presence of pore fluid. We introduce an effective density ratio $\mathcal{D} = \frac{\rho_h - \rho_f}{\rho_l - \rho_f}$ to characterise this combined effect. According to our numerical results, for $\mathcal{D} < 5$, the system remains at a well-mixed stage, whereas for $\mathcal{D} \geq 5$, nearly complete segregation occurs. The dashed diagonal line in Fig. 2(a) marks the transition states corresponding to $\mathcal{D} = \frac{\rho_h - \rho_f}{\rho_l - \rho_f} = 5$. However, this transition threshold is arbitrary, as the transition is gradual and not strictly defined by $\mathcal{D} = 5$; nonetheless, it is evident that $M$ is strongly influenced by $\mathcal{D}$.

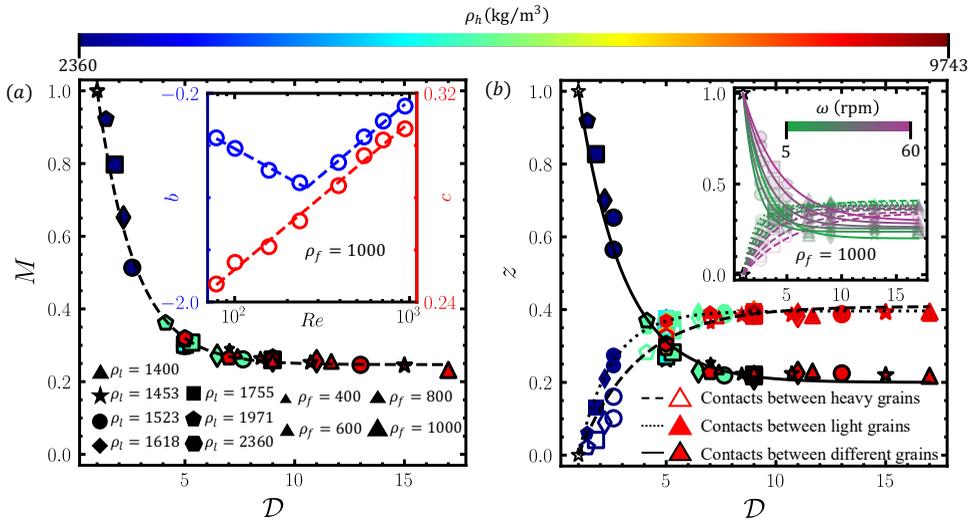

FIGURE 3. (a) Mixing index $M$ as a function for the effective density ratio $\mathcal{D}$ at $\omega = 5$ rpm. Dashed lines indicate the fitted functions, and the hollow pentagram denotes $(\mathcal{D}, M) = (1,1)$. The inset illustrates the parameters, $b$ and $c$ vs. $Re$. (b) The contact number ratio, $z$, plotted against $\mathcal{D}$ at $\omega = 5$ rpm. The upper hollow pentagram is for $(\mathcal{D}, z) = (1,1)$, while the bottom hollow pentagram is for $(\mathcal{D}, z) = (1,0)$. The inset is for $z$ versus $\mathcal{D}$ for various rotating speeds.

As demonstrated in Fig. 3(a), wherein all numerical realisations in Fig. 2(a) are included, $M$ monotonically decreases with $\mathcal{D}$ following an exponential law, $M = (1-c)e^{b(\mathcal{D}-1)} + c$, with two parameters $b < 0$, and $0 < c < 1$. Considering two extreme cases: mono-disperse grain mixtures, for $\rho_h = \rho_l$ we have $(\mathcal{D}, M) = (1,1)$; and for $\rho_l \to \rho_f$, $\lim_{\rho_l \to \rho_f} \mathcal{D} = +\infty \to \lim_{\rho_l \to \rho_f} M = c$, i.e., $(\mathcal{D}, M) = (+\infty, c)$. The parameter $c$ sets the segregation limit in our binary mixture system, and the actual value depends on the drum conditions, including the geometry, speed, and filling state. Specific evolutions of $b$ and $c$ with the global Reynolds number, $Re_g$ are provided in the inset of Fig. 3(a). Interestingly, $c$ is logarithmically linear to $Re_g$, while $b$ is logarithmically bilinear to $Re_g$, indicating the wet granular flow transition from the viscous- to inertial-dependent regime (Topin et al., 2012).

Since $M$ is rooted in contact numbers—readily accessible in simulations—of the associated mixtures, to gain deeper insights into how it evolves with $\mathcal{D}$, Fig. 3(b) shows the proportions ($z_h$, $z_l$, and $z_d$) in the total contact number of three contact types, namely contacts between heavy grains, contact between light grains, and contacts between different types of grains. The contact is determined by a distance criterion between any two adjacent particles whose centre distance, $l \leq 1.05 d_p$. This criterion is also implemented for the validations in the inset of Fig. 2(a). Interestingly, all these three proportions evolve with varying $\mathcal{D}$ in an exponential fashion like $M$, indicating that the exponential dependence of $M$ on $\mathcal{D}$ can be explained by contact proportions evolutions. Therein, at $\mathcal{D} = 1$, the submerged D-system degrades to mono-disperse wet mixtures and, thus the two types of particles are perfectly mixed as long as rotating revolutions are enough. This could induce no contact between different types of grains; that is, $z_h = z_l = 0$ and $z_d = 1$. With increased $\mathcal{D}$ while keeping $\omega$ consistent, $z_h$ and $z_l$ rise and gradually saturate because of the emergence of complete segregations. Moreover, akin to $M$, the $\omega$-dependent exponential evolutions with $\mathcal{D}$ are also demonstrated in the inset of Fig. 3(b). When the drum is rotated more rapidly, the wider spatial allocation of both types of grains results in more area of the contact boundary between the heavy grain clusters and its surrounding light grains, as illustrated in the top row of Fig. 1(b). This explains why higher $\omega$ contributes to larger $z_d$. Meanwhile, $z_h$ and $z_l$ are decreased, since $z_h + z_l + z_d = 1$.

### 3.2 *Segregation pattern transitions*

From the three experimentally observed segregation patterns in Fig. 1(b)—each exhibiting one, two, or three vortices (the third, typically smaller, forming near the top centre of the mixture)—all the mixing indices calculated from the corresponding simulation are less than 0.29, but the mixing patterns are totally different. To quantitatively describe such a difference in the nearly complete segregation patterns, we measure the vortex information in them. With both increases of $\mathcal{D}$ and $\omega$, the trajectories depicted in the bottom row of Fig. 1(b) reveal a transition from a single-core pattern to two- or three-cores pattern, with the counterclockwise motion of the heavy grain cluster parallel to the drum rotating direction. It has been widely observed that in a dry D-system heavy grains could congregate in the core of the symmetrical mixture (Hill et al., 1999). As a thought experiment, fluids are assumed to be added to such a system in a steady state; then a macro force, combined with counter-gravity buoyance and drag forces with net parallelising to rotating velocity, on the heavy grain cluster could move it to the concurrent position and induce the asymmetry. This asymmetry could be enhanced with rising $\omega$ that induces higher drag forces as described in Eq. (5). Additionally, at a higher $\mathcal{D}$, light grains behave similarly to the surrounding fluid, allowing heavy grains to exhibit independent motions—like a system just for the mixture of heavy grains and fluids—and to get closer to the cylinder boundary.

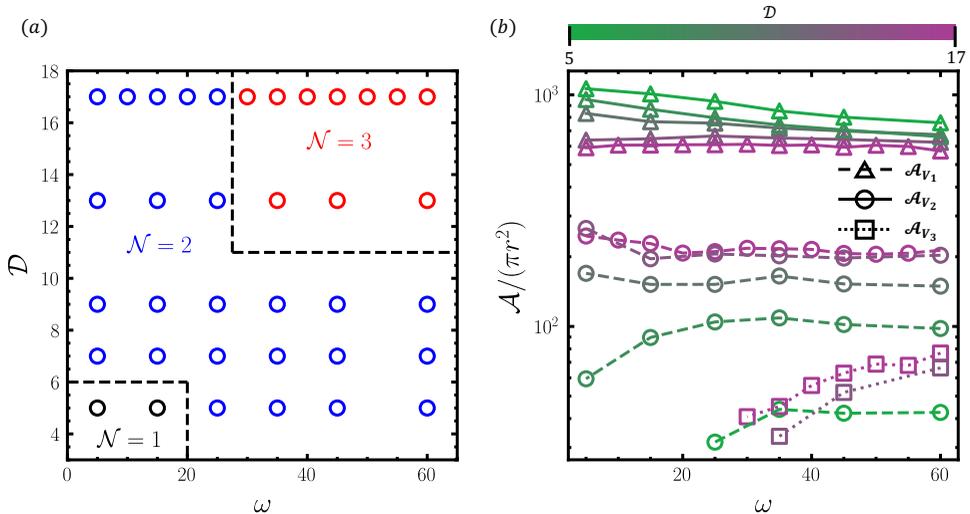

FIGURE 4. (a) Phase diagram of the measured number of vortices, $\mathcal{N}$, influenced by the rotating speed, $\omega$, and the effective density ratio, $\mathcal{D}$, as also shown in Fig. 1(b). (b) The normalized vortex area, $\mathcal{A}$, by projected area of single grain versus $\omega$, with different $\mathcal{D}$.

Influences of $\mathcal{D}$ and $\omega$ on modes of the nearly complete segregation are detangled via vortex number, $\mathcal{N}$, and vortex area, $\mathcal{A}$. To this end, the codes (Endrikat, 2024) in the MATLAB® environment for $\Gamma_2$ criterion proposed by Graftieaux et al. (2001) are implemented to identify and measure vortexes. All illustrative visualisations of three identified vortex configurations are provided in Fig. 1(b). From the phase diagram in Fig. 4(a), dry D-system like patterns with one vortex ($V_1$) in heavy grains could only attain at low values of both $\mathcal{D}$ and $\omega$. Amplifying $\mathcal{D}$ or $\omega$ can induce a second vortex ($V_2$) in light grains. A third vortex ($V_3$), located near the centre of the mixture in light grains, could emerge at both high $\mathcal{D}$ and $\omega$. Fig. 4(b) shows the combined effect of $\mathcal{D}$ and $\omega$ on evolutions of the three vortexes areas. For $V_1$, the vortex area persists regardless of $\omega$ at higher $\mathcal{D}$; however, due to the relatively less inertial forces of heavy grains to light grains at lower $\mathcal{D}$, the persistency of $\mathcal{A}_{V_1}$ with varying $\omega$ is diminished. This tendency also suits the evolutions of $\mathcal{A}_{V_2}$. Compared with $\mathcal{A}_{V_1}$, $\mathcal{A}_{V_2}$ reduces rapidly when $\mathcal{D}$ is decreased, because the movement of lighter grains can fluctuate more and deviate from the stable trace lines forming the vortex. Interestingly, at the extremely low $\mathcal{D} = 5$, $\mathcal{A}_{V_2}$ could be even smaller than $\mathcal{A}_{V_3}$, considering generally we have $\mathcal{A}_{V_1} > \mathcal{A}_{V_2} > \mathcal{A}_{V_3}$.

3.3 *Discussions*

The results demonstrate that the CFD-DEM approach accurately reproduces the experimentally observed segregation patterns. At a fixed $\omega$, the mixing efficiency, quantified by the Lacey index ($M$), is controlled primarily by $\mathcal{D}$. This dependence is well-described by an exponential law in which the scaling coefficient $c$ increases monotonically with $Re_g$, while the $M - \mathcal{D}$ rate factor, $b$, transitions from decreasing to increasing at a critical $Re_g$. Along with these changes in the global scalar fields, the flow undergoes spatially heterogeneous topological bifurcations, marked by the emergence of multiple coherent vortical structures. Specifically, at low $\omega$ and small $\mathcal{D}$, the system exhibits a single focal point of species segregation and drum-relative circulation; however, higher rotation rates or larger density contrasts generate a second core domain in which lighter grains collect and circulate, and a third vortex can eventually form at even higher density contrasts.

Although the current work systematically explores the effects of $\mathcal{D}$ and $\omega$, the influence of the drum filling fraction—known to induce quasi-static regions in dry systems (Maguire et al., 2024)—has yet to be addressed under submerged conditions, nor has the impact of drum geometry (e.g. circular, elliptical, or polygonal) been fully explored. Moreover, fully characterising the bilinear change in the coefficient $b$ and clarifying the viscous-inertial transition at the particle scale—particularly the role of the interstitial fluid viscosity—remains essential (Cui et al., 2022; Trulsson et al., 2012). For instance, it warrants further investigation of linking these transitions to the local Stokes number or other comparable grain-scale metrics. While global measures such as the Lacey index effectively capture overall mixing states, coupling local flow rheology and stress-segregation can potentially pave the way towards enriching and combining the existing scaling theories, such as the submerged grainsize (Trewhela et al., 2021) and density segregation frameworks (Gray and Ancey, 2015).

## 4. Conclusion

This study provides a comprehensive analysis of density-driven segregation of binary granular mixtures in a rotating drum. The observations from experiments and numerical simulations reveal four distinct mixing states—from well mixed states to complicated segregation patterns with increasing vortices. A new factor, effective density ratio, $\mathcal{D}$, is defined to bridge the grain and fluid properties and the macroscopic mixing index, $M$. A new model is also proposed to predict $M$ of a D-system. The nearly complete segregation patterns transformed from the well mixing state is classified by quantitative analysis of vortices in the two grain phases. The vortex area and number are highly sensitive to $\mathcal{D}$ and $\omega$. These results extend our understanding of submerged grain segregation, providing critical insights into how to segregate two mixed dry grains by adding the specific fluids. The identification of the scaling laws between $M$ and $\mathcal{D}$ contributes to broader science and engineering communities of granular flow, such as geophysics and industrial manufacturing.

**Acknowledgements.** DW acknowledges the financial supports by the National Natural Science Foundation of China (the Excellent Young Scientists Fund (overseas)) and by the Australian Research Council (ARC, the Discovery Early Career Award (DECRA), Grant No. DE240101106).

**Declaration of interest.** The authors report no conflict of interest.

**Author contributions. Yu Chen:** Conceptualisation, Formal analysis, Methodology, Visualisation, Writing – original draft, Writing – review & editing; **Deheng Wei:** Conceptualisation, Formal analysis, Methodology, Visualisation, Supervision, Writing – original draft, Writing – review & editing; **Si Suo:** Methodology, Writing – review & editing; **Mingrui Dong:** Writing – review & editing; **Yixiang Gan:** Conceptualisation, Formal analysis, Methodology, Visualisation, Supervision, Writing – review & editing.